\documentclass[sigplan,screen]{acmart}

\copyrightyear{2022}
\acmYear{2022}
\setcopyright{acmcopyright}\acmConference[ICCAD '22]{IEEE/ACM International Conference on Computer-Aided Design}{October 30-November 3, 2022}{San Diego, CA, USA}
\acmBooktitle{IEEE/ACM International Conference on Computer-Aided Design (ICCAD '22), October 30-November 3, 2022, San Diego, CA, USA}
\acmPrice{15.00}
\acmDOI{10.1145/3508352.3549478}
\acmISBN{978-1-4503-9217-4/22/10}

\usepackage{amsmath,amssymb,amsfonts}
\usepackage{graphicx}
\usepackage{algorithm}
\usepackage{algpseudocode}
\usepackage{threeparttable}
\usepackage{graphics}
\usepackage{epsfig}
\usepackage{textcomp}
\usepackage{xcolor}
\usepackage{amssymb}
\usepackage{booktabs}
\usepackage{multirow}
\usepackage{makecell}

\usepackage{hyperref}
\usepackage{url}
\usepackage{mathtools}

\AtBeginDocument{%
  \providecommand\BibTeX{{%
    \normalfont B\kern-0.5em{\scshape i\kern-0.25em b}\kern-0.8em\TeX}}}

\definecolor{Note_color}{rgb}{1.0, 0.0, 0.0}

\newcommand\blfootnote[1]{%
  \begingroup
  \renewcommand\thefootnote{}\footnote{#1}%
  \addtocounter{footnote}{-1}%
  \endgroup
}

\usepackage{enumitem}
\setlist[itemize]{leftmargin=0.5cm}

\title{NASA: Neural Architecture Search and Acceleration for Hardware Inspired Hybrid Networks}

\author{Huihong Shi$^{1,2*}$, Haoran You$^1$, Yang Zhao$^3$, Zhongfeng Wang$^{2}$, and Yingyan Lin$^1$}

\affiliation{
\institution{
$^1$Georgia Institute of Technology,
$^2$Nanjing University,
$^3$Rice University, }
\country{$^1$USA, $^2$P.R. China, $^3$USA,}
}

\email{
{eiclab, hyou37, celine.lin}@gatech.edu,
zy34@rice.edu, 
zfwang@nju.edu.cn
}

\def \authors{Huihong Shi, Haoran You, Yang Zhao, Zhongfeng Wang, Yingyan Lin}
\renewcommand{\shortauthors}{Huihong Shi, Haoran You, Yang Zhao, Zhongfeng Wang, Yingyan Lin}
\renewcommand{\shorttitle}{Neural Architecture Search and Acceleration for Hardware Inspired Hybrid Networks}

\begin{document}

\begin{abstract}
Multiplication is arguably the most cost-dominant operation in modern deep neural networks (DNNs), limiting their achievable efficiency and thus more extensive deployment \sloppy in resource-constrained applications. To tackle this limitation, pioneering works have developed handcrafted multiplication-free DNNs, which require expert knowledge and time-consuming manual iteration, calling for fast development tools. To this end, we propose a \textbf{N}eural \textbf{A}rchitecture \textbf{S}earch and \textbf{A}cceleration framework dubbed NASA, which enables automated multiplication-reduced DNN development and integrates a dedicated multiplication-reduced accelerator for boosting DNNs' achievable efficiency. Specifically, NASA adopts neural architecture search (NAS) spaces that augment the state-of-the-art one with hardware inspired multiplication-free operators, such as shift and adder, armed with a novel progressive pretrain strategy (PGP) together with customized training recipes to automatically search for optimal multiplication-reduced DNNs; On top of that, NASA further develops a dedicated accelerator, which advocates a chunk-based template and auto-mapper dedicated for NASA-NAS resulting DNNs to better leverage their algorithmic properties for boosting hardware efficiency. Experimental results and ablation studies consistently validate the advantages of NASA's algorithm-hardware co-design framework in terms of achievable accuracy and efficiency tradeoffs. 
Codes are available at  {\url{https://github.com/GATECH-EIC/NASA}}.

\blfootnote{\textsuperscript{$*$}
Work done when Huihong was a visiting student at Georgia Tech.
Correspondence should be addressed to: Zhongfeng Wang and Yingyan Lin.}

\end{abstract}
\maketitle


\vspace{-1em}
\section{Introduction}
\vspace{-0.2em}
Modern deep neural networks (DNNs) have achieved great success in various computer vision tasks  \cite{He2016DeepRL, GoingDW, YOLO9000BF, YOLOv3AI}, which has motivated a substantially increased demand for DNN-powered solutions in numerous real-world applications. However, the extensively used multiplications in DNNs dominate their energy consumption and have largely challenged DNNs’ achievable hardware efficiency, motivating \sloppy multiplication-free DNNs that adopt hardware-friendly operators, such as additions and bit-wise shifts, which require a smaller unit energy and area cost as compared to multiplications \cite{ShiftAddNet}.
In particular, pioneering works of multiplication-free DNNs include (1)
DeepShift \cite{DeepShift} which proposes to adopt merely shift layers for DNNs, (2)
AdderNet \cite{AdderNet} which advocates using adder layers to implement DNNs for trading the massive multiplications with lower-cost additions, and (3) ShiftAddNet \cite{ShiftAddNet} which combines both shift and adder layers to construct DNNs for better trading-off the achievable accuracy and efficiency.

Despite the promising hardware efficiency of the multiplication-free DNNs, their models' expressiveness capacity and thus achievable accuracy are generally inferior to their multiplication-based counterparts. As such, it is highly desired to develop hybrid multiplication-reduced DNNs that integrate both multiplication-based and multiplication-free operators (e.g., shift and adder) to boost the hardware efficiency while maintaining the task accuracy.
Motivated by the recent success of neural architecture search (NAS) in automating the design of efficient and accurate DNNs, one natural thought is to leverage NAS to automatically search for the aforementioned hybrid DNNs for various applications and tasks, each of which often requires a different accuracy-efficiency trade-off and thus calls for a dedicated design of the algorithms and their corresponding accelerators.
In parallel, various techniques \cite{Eyeriss, FusedlayerCA, SmartExchange, DNNBuilderAA, MaximizingCA, AdderNetAI} have been proposed to boost the hardware efficiency of DNNs, promoting their real-world deployment from the hardware perspective.
For example, Eyeriss \cite{Eyeriss} proposes a row stationary dataflow and a micro-architecture with hierarchical memories to enhance data locality and minimize the dominant data movement cost; and \cite{AdderNetAI} explores a low-bit quantization algorithm paired with a minimalist hardware design for AdderNet \cite{AdderNet} to leverage its algorithmic benefits for boosted hardware efficiency.                                      
While it has been shown that dedicated accelerators can achieve up to three orders-of-magnitude efficiency improvement as compared to general computing platforms, such as GPUs and CPUs, existing accelerators are customized for either multiplication-based or multiplication-free DNNs, and thus could not fully leverage the algorithmic properties of the aforementioned hybrid DNNs for maximal efficiency. Thus, it is promising and desirable to develop dedicated accelerators for hybrid DNNs consisting of both multiplication-based and multiplication-free operators, which yet is still underexplored.

\begin{figure}[t]
	\centerline{\includegraphics[width=\linewidth]{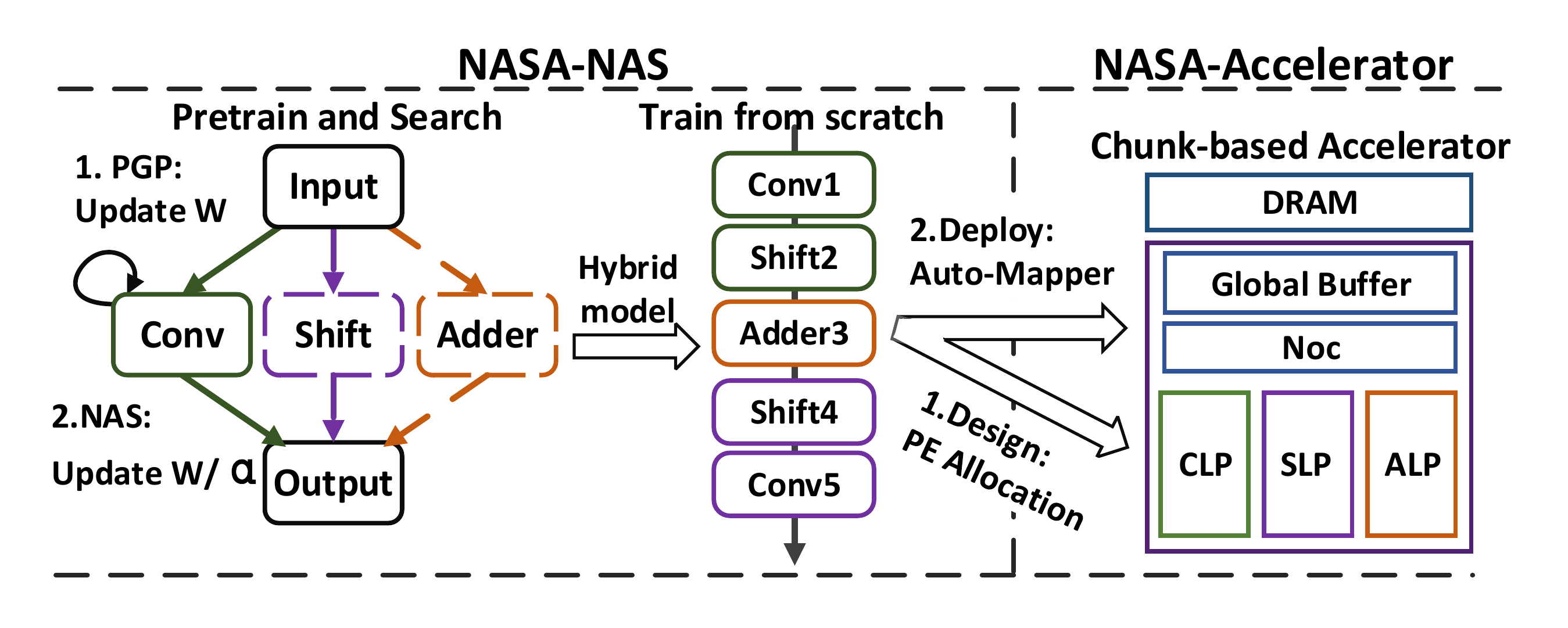}} \vspace{-1.5em}
	\caption{\small An overview of our NASA framework integrating neural architecture search (NAS) and acceleration engines dedicated for hybrid DNNs, where NASA-NAS searches for hybrid models via NAS with the proposed progressive pretrain strategy (PGP) while NASA-Accelerator advocates a dedicated chunk-based design armed with an auto-mapper to support NASA-NAS searched DNNs. } \vspace{-2.1em}
	\label{fig:framework} 
\end{figure}

To marry the best of both worlds - the higher achievable accuracy of multiplication-based DNNs and the better hardware efficiency of multiplication-free DNNs, we target the \emph{\textbf{exploration} and \textbf{acceleration} of \textbf{hybrid DNNs}}, and make the following contributions: 

\begin{itemize}
   \item We propose \textbf{NASA}, a \textbf{N}eural \textbf{A}rchitecture \textbf{S}earch and \sloppy \textbf{A}cceleration framework (see Fig \ref{fig:framework}) to search for and accelerate hardware inspired hybrid DNNs. To the best of our knowledge, NASA is \textbf{\emph{the first}} algorithm and hardware co-design framework dedicated for hybrid DNNs.
    \item We develop a dedicated NAS engine called NASA-NAS integrated in NASA, which incorporates hardware-friendly shift layers \cite{DeepShift} and/or adder layers \cite{AdderNet} into a state-of-the-art (SOTA) hardware friendly NAS search space \cite{FBNet} to construct \emph{hybrid DNN search spaces}. Furthermore, to enable effective NAS on top of the hybrid search space, we propose a \emph{\textbf{P}ro\textbf{G}ressive \textbf{P}retrain strategy (PGP)} paired with \emph{customized training recipes}.
    \item We further develop a dedicated accelerator called NASA-Accelerator to better leverage the algorithmic properties of hybrid DNNs for improved hardware efficiency. Our NASA-Accelerator advocates \emph{a dedicated chunk-based accelerator} to better support the heterogeneous layers in hybrid DNNs, and integrates \emph{an auto-mapper} to automatically search for optimal dataflows for executing hybrid DNNs in the above chunk-based accelerators to further improve efficiency. 
    \vspace{0.15em}
    \item Extensive experiments and ablation studies validate the effectiveness and advantages of our NASA in terms of achievable accuracy and efficiency tradeoffs, against both SOTA multiplication-free and multiplication-based systems. We believe our work can open up an exiting perspective for the exploration and deployment of multiplication-reduced hybrid models to boost both task accuracy and efficiency.
\end{itemize}

\vspace{-2em}
\section{Related Works}
\subsection{Multiplication-free DNNs}
To favor hardware efficiency, pioneering efforts have been made to replace the cost-dominant multiplications in vanilla DNNs with more hardware-friendly operators, e.g., bit-wise shift and adder, for enabling handcrafted multiplication-free DNNs.
For instance, ShiftNet \cite{shift} treats shift operations as a zero flop/parameter alternative to spatial convolutions and advocates DNNs featuring shift layers; 
DeepShift \cite{DeepShift} substitutes multiplications with bit-wise shifts; 
AdderNets \cite{ AdderNet} trades multiplications with lower-cost additions and employs an $\ell$1-normal distance as a cross-correlation substitute to measure the similarity between input features and weights; and inspired by a common hardware practice that implements multiplications with logical bitwise shifts and additions \cite{AdaptiveEU},
ShiftAddNet \cite{ShiftAddNet} unifies both shift and adder layers to design DNNs with merely shift and adder operators.  
However, multiplication-free DNNs in general are still inferior to their multiplication-based counterparts in terms of task accuracy, motivating our NASA framework aiming to marry the best of both worlds from powerful multiplication-based and hardware efficient multiplication-free DNNs.  
\vspace{-1.2em}
\subsection{Neural Architecture Search}
\vspace{-0.2em}
Early NAS methods \cite{NAS, ZophNAS, RegularizedEF} utilize reinforcement learning (RL) to search for DNN architectures, which gained great success but were resource- and time-consuming. 
To tackle this limitation, weight sharing methods \cite{EfficientNAS, darts, ProxylessNAS} have been proposed.
Among them, differentiable NAS (DNAS) algorithms \cite{darts, FBNet, ProxylessNAS} have achieved SOTA results by relaxing the discrete search space to be continuous and then
applying gradient-based optimization methods to find optimal architectures from a pre-defined differentiable supernet.
Specifically, FBNet \cite{FBNet} employs a Gumbel Softmax sampling method \cite{GS} and gradient-based optimization to search for efficient and accurate DNNs targeting mobile devices;  
FBNetV2 \cite{FBNetV2} designs a masking mechanism for feature map reuses in both spatial and channel dimensions, expanding the search space greatly at a cost of a small memory overhead; 
alternatively, ProxylessNAS \cite{ProxylessNAS} activates only a few paths during the forward and backward processes of search, making it possible for DNAS to optimize with large search spaces. 
Despite the prosperity of NAS for vanilla DNNs, there still lacks efforts in exploring NAS designs for hybrid DNNs.
\vspace{-1.0em}
\subsection{DNN Accelerators}
\vspace{-0.3em}
Apart from algorithmic optimization, many works \cite{Eyeriss,FusedlayerCA,SmartExchange,DNNBuilderAA,AdderNetAI,MaximizingCA} have proposed dedicated DNN accelerators to boost DNNs' hardware efficiency. 
Among them, Eyeriss \cite{Eyeriss} proposes a row stationary dataflow and a micro-architecture with hierarchical memories to enhance data locality and minimize dominant data movement costs;
\cite{10.1145/3373087.3375306} proposes AutoDNNchip, a DNN chip generator that can automatically generate both FPGA- and ASIC-based DNN chip implementation given DNNs from machine learning frameworks for a designated application and dataset; and
\cite{MaximizingCA} builds a DNN accelerator to increase the overall acceleration throughput by partitioning the FPGA resources into multiple specialized processors for various convolution layers. 
Nevertheless, existing mainstream accelerators are dedicated for either  multiplication-based DNNs or multiplication-free DNNs (e.g., \cite{AdderNetAI} for AdderNet \cite{AdderNet}), 
which thus can not fully leverage our target hybrid DNNs' hardware efficiency, 
calling for dedicated accelerators to unleash hybrid DNNs' full efficiency potential.
\vspace{-1.0em}
\section{Proposed NASA-NAS Engine} \label{NASA-NAS}
In this section, we first introduce our proposed search spaces that integrates both multiplication-free operators and multiplication-based convolutions, and 
then present our progressive pretrain strategy (PGP) proposed to enable effective training of hybrid supernets (see Sec. \ref{sec:progressive_training}). 
Finally, Sec. \ref{sec:search_alg} describes the adopted NAS algorithm based on PGP to search for hardware-inspired hybrid DNNs.

\vspace{-0.5em}   
\subsection{NASA-NAS's Search Spaces}
\label{sec:search_space}
\textbf{Basic operators.} To push forward the frontier of achievable task accuracy and efficiency tradeoffs, our NASA-NAS search space unifies both hardware-friendly multiplication-free operators, such as coarse-grained \emph{shift layers} \cite{DeepShift} and fine-grained \emph{adder layers} \cite{AdderNet}, and multiplication-based \emph{convolutions} to construct hybrid search spaces. Next, we will introduce the basic shift and adder layers.

\begin{figure}[!t]
	\centerline{\includegraphics[width=3.5in,height=0.8in]{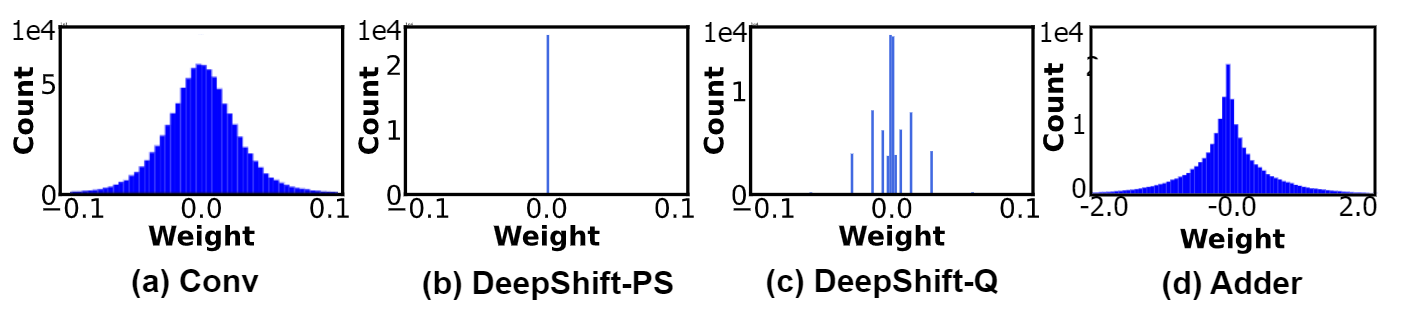}}
	\vspace{-1.2em}
	\caption{Illustrating the weight distributions of (a) convolutions, shift layers constructed with (b) DeepShift-PS and (c) DeepShift-Q, respectively, and (d) adder layers, from a NASA-NAS searched hybrid-all model on CIFAR100.}
	\vspace{-1.5em}
	\label{fig:shift_distribution}
\end{figure}

\begin{itemize}
\item \textbf{Shift layers.} 
    The promising hardware efficiency of bit-wise shift operations has motivated implementing DNNs with shift layers \cite{DeepShift} as formulated in Eq. (\ref{eq:shift}), where the weight $W_{shift}$ can be constructed through two ways: \emph{DeepShift-PS} and \emph{DeepShift-Q}. 
    As formulated in Eq. (\ref{eq:deppshift-ps}), DeepShift-PS intuitively parametrizes $W_{shift}$ with sign flipping ${s}$ and bit-wise shift ${p}$, in which $s\in[-1,0,1]$ and ${p}$ is an integer with an absolute value normally greater than one. Given the example that $W_{shift}= -0.25$/$-0.125$, then the parameter $s=-1$ and $p=-2$/$-3$, and thus we can easily conclude that the smaller $W_{shift}$ comes with the larger absolute value of $p$. As such, parameters of shift layers that have larger absolute values are naturally incompatible with the small weights in convolutions (see Fig. \ref{fig:shift_distribution}a), yielding $W_{shift}$ of shift layers in hybrid DNNs to be zero (see Fig. \ref{fig:shift_distribution}b). 
    
    \quad To tackle the issue above, we leverage a more training friendly way dubbed \emph{DeepShift-Q} to build our hybrid DNNs, which generates $W_{shift}$ by quantizing the vanilla weights in convolutions $w^{*}$ to power of two following Eq. (\ref{eq:shift-q}) instead of directly optimizing $p$ and $s$, and Fig. \ref{fig:shift_distribution}c demonstrates its effectiveness.
    \begin{equation}
		Y = \sum X^\mathrm{T}*W_{shift}. \label{eq:shift}
    \end{equation}
    \vspace{-1em}
    \begin{equation}
        W_{shift} = s*2^{p}.
    \label{eq:deppshift-ps}
    \end{equation}
    \vspace{-1em}
    \begin{equation}
		\quad \hat{s}=sign(w^{*}), \ \hat{p}=round(log_{2}|w^{*}|),
		\ w_{shift}=\hat{s}*2^{\hat{p}}.
		\label{eq:shift-q}
    \end{equation}
    
    \item \textbf{Adder layers.}
    The multiplication-intensive convolutions in DNNs are essentially cross-correlation functions used to measure the relevance between activations and weights.
    Alternatively, AdderNet \cite{AdderNet} proposes adder layers, which leverage computational efficient additions and calculate the $\ell1$-norm distance to measure the relevance between activations and weights, as expressed in Eq. (\ref{eq:adder}). 
    Due to the different properties of adder layers and convolutions, existing pretrain and NAS algorithms are not directly applicable to handle our hybrid-adder search space, motivating our proposed \emph{progressive pretrain strategy} introduced in Sec. \ref{sec:progressive_training}
    \begin{equation}
		Y =\sum -\left\lvert X- W_{adder}\right\rvert. 
		\label{eq:adder}
    \end{equation} 
    \end{itemize}

Based on the above shift and adder layers, we construct three hybrid search spaces: \textbf{\emph{hybrid-shift}}, \textbf{\emph{hybrid-adder}}, and \textbf{\emph{hybrid-all}}, as summarized in Table \ref{table:choice}, by integrating them separately or jointly into the SOTA convolutional one \cite{FBNet}.

\begin{figure}[!t]
	\vspace{-0.2em}
	\centerline{\includegraphics[width=2.3in,height=1.5in]{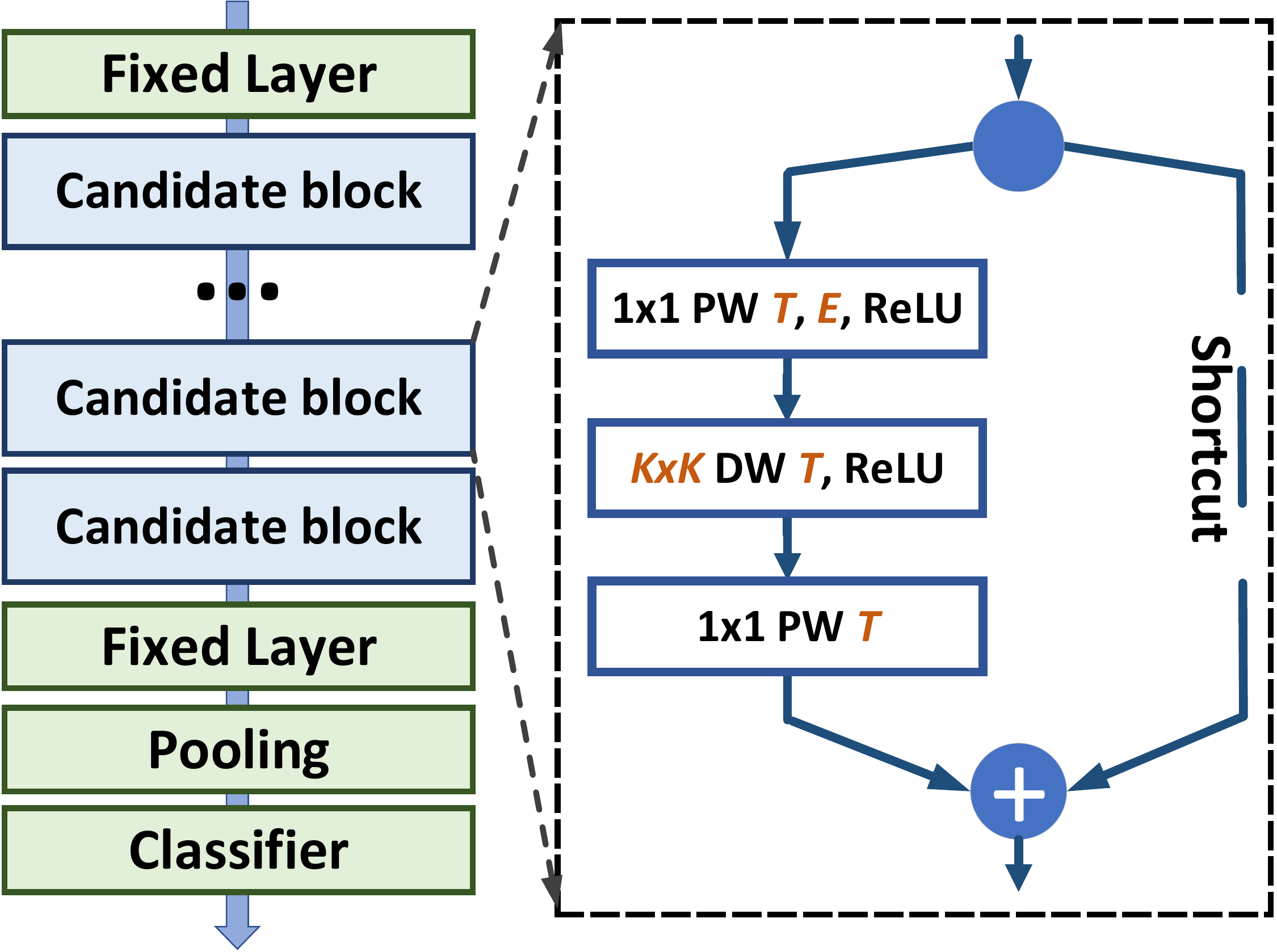}}
	\vspace{-1.5em}
\caption{An overview of our supernet, where both the macro-architecture (left) and the detailed candidate blocks (right) are shown. Here PW, and DW denote pointwise and depthwise layers, and $E, K$, and $T$ donate the channel expansion ratio of candidate blocks, kernel size in depthwise layers, and layer type, respectively.}
	\label{fig:supernet}
	\vspace{-1.8em}
\end{figure}

\textbf{Supernet.} Built upon the proposed search spaces above, we construct our supernet in Fig. \ref{fig:supernet}. 
As depicted in Fig. \ref{fig:supernet} (left), we adopt a macro-architecture following \cite{FBNet} for ease of benchmarking with SOTA efficient DNNs like \cite{FBNet}, where the first and last three layers are fixed and the rest candidate blocks are searchable from our pre-defined search spaces. 
As shown in Fig. \ref{fig:supernet} (right), each candidate block consists of two pointwise layers (PW) and one depthwise layer (DW), and is characterized by three hyperparameters: the channel expansion ratio of the block $E$, kernel size of the DW layer $K$, and layer type $T$.
The configurations of these hyperparameters are summarized in Table \ref{table:choice}: All three search spaces contain the same options for $E$ and $K$ but \emph{different choices for $T$}, e.g., the hybrid-shift/adder search space integrates convolutions with only shift/adder layers, while the hybrid-all one includes both of them; Additionally, a \emph{skip} operator is included to skip unnecessary blocks and thus allow shallower DNNs. It is worth noting that we adopt shared weights 
between candidate blocks with the same layer type $T$ and kernel size $K$ among the channel dimension $E$ to enable effective NAS, inspired by \cite{HAT}. 

\begin{table}[]
	\centering
	\caption{Configurations of candidate blocks in our pre-defined search spaces, where $E, K, T, $ and Skip donate the expansion ratio, kernel size, layer type, and skip operators, respectively, and Conv, Shift, Adder represent convolutions, shift layers,  and adder layers.}
	\vspace{-1em}
	\resizebox{0.9\linewidth}{!}{
	\begin{tabular}{c|c|c}
	\hline
	Hyperparameters           & Choices                   & Search Spaces       \\ \hline
	\multirow{2}{*}{\begin{tabular}[c]{@{}c@{}}($E$, $K$)\end{tabular}} & (1,3),(3,3),(6,3)      & \multirow{2}{*}{All Search Spaces} \\         
				& (1,5),(3,5),(6,5)                      &          \\ \hline
	\multirow{3}{*}{$T$}    & Conv, Shift                & Hybrid-Shift     \\
				& Conv, Adder                 & Hybrid-Adder       \\
				& Conv, Shift, Adder          & Hybrid-All    \\ \hline
	Skip        & --                          &  All Search Spaces               \\ \hline																					
	\end{tabular}} \label{table:choice} \vspace{-2.0em}
	\end{table} 

In summary, there are a total of $22$ layers to be searched in our supernet as \cite{FBNet}, each of them can select from {$13$} or $19$ (corresponding to the num. of ($E,K$) $ \times$ the num. of $T$ + 1) candidate blocks from Table \ref{table:choice} to construct the hybrid-shift/hybrid-adder or hybrid-all search spaces, respectively.
For instance, there are (1) $6$ choices for ($E,K$), pairing the expansion ratio $E$ and kernel size $K$, (2) $2$ (or $3$) choices for the layer type $T$, and (3) $1$ skip operation in the hybrid-shift/adder (or hybrid-all) search space based on Table \ref{table:choice}; As such, each searchable layer in our hybrid-shift/adder (or hybrid-all) supernet can choose from $6\times2 +1=13$ (or $6\times3 +1=19$) candidate blocks.
Hence, there are a total of $13^{22}$ or $19^{22}$ potential architectures in our pre-defined search spaces.

\vspace{-0.8em}
\subsection{NASA-NAS's Progressive Pretrain Strategy}
\label{sec:progressive_training}

We empirically find that directly applying the SOTA NAS pretrain and search method \cite{FBNet} to our \emph{hybrid-adder} search space consistently suffers from the non-convergence issue, and identify that this is caused by the distinct weight distributions of adder and convolutional layers (denoted as conv). Specifically, weights of adder layers tend to follow a Laplacian weight distribution (see Fig. \ref{fig:shift_distribution}d), while those of convolutional layers normally follow a Gaussian distribution (see Fig. \ref{fig:shift_distribution}a).     
To tackle this issue, we propose a \emph{progressive pretrain strategy (PGP)} that can enable effective pretraining of our target hybrid-adder supernets through \emph{three stages}: (1) conv pretraining, (2) adder pretraining with fixed conv, and (3) mixture pretraining, paired with our \emph{customized training recipe} (i.e., bigger learning rate (lr) and initialization). 

Next, we elaborate the three stages above. 
Specifically, in the first (1) conv pretraining stage, we only forward and backward the convolution-based blocks to first leverage the high-speed convergence benefit of vanilla CNNs, providing a good initialization for the target hybrid supernet; in stage (2), we freeze the well-pretrained weights in convolutional layers to facilitate training the adder layers, in which we forward both convolutional and adder layers but only backward the latter during the backward process; and finally in the (3) mixture pretraining stage, we free up the previously fixed convolutional layers and simultaneously optimize all the candidate blocks to coordinate their parameters towards optimal model accuracy. 
As for the customized \emph{training recipe}, we empirically observe that the lr can greatly affect the optimization of hybrid-adder supernets, and a \emph{bigger lr} can accelerate the convergence; Additionally,
we find that setting the learnable scaling coefficient $\gamma$ in the last batch normalization layer of each candidate block to $0$, similar to \cite{BigNAS}, favors effective training of hybrid supernets.

Note that while we introduce the non-convergence issue above using the \emph{hybrid-adder} search space, it also exists in the \emph{hybrid-all} search space where our PGP strategy is equally effective. On the other hand, for the \emph{hybrid-shift} search space, the vanilla pretrain method in \cite{FBNet} is sufficient. More details can be found in Sec. \ref{sec:exp_alg}.   

\vspace{-0.5em}
\subsection{NASA-NAS's Search Algorithm}
\label{sec:search_alg}

In NASA-NAS, we adopt SOTA differentiable NAS (DNAS) algorithm \cite{FBNet}, which relaxes the discrete search space to be continuous and then applies gradient-based optimization to optimize the weights $w$ and architectural parameters $\alpha$ in pre-defined differentiable supernets, to enable effective search for our target hybrid models. 
Specifically, we update weights $w$ in both hybrid-adder and hybrid-all supernets on top of the proposed \emph{PGP} strategy, which alleviates the non-convergence issue via progressively optimizing heterogeneous layers and thus promoting the integration of adder and other operators, such as convolutions and shift layers. 
The loss function is formulated as:
\vspace{-0.3em}
\begin{equation}
\small
	\mathop{min}\limits_{\alpha}\
	\mathcal{L}_{ce}^{val}(\alpha,w*)+
	\lambda\mathcal{L}_{hw}(\alpha), 
	\\ \mathop{s.t.} \ w*=\mathop{min}\limits_{w}\mathcal{L}_{ce}^{train}(\alpha,w),
	\label{loss} \vspace{-0.6em}
\end{equation} 
where $\mathcal{L}_{ce}^{train}$ and $\mathcal{L}_{ce}^{val}$ denote the cross-entropy loss on the training and validation datasets, respectively. 
We apply FLOPs (floating-point operations per second) as a proxy hardware metric to calculate the hardware-aware loss $\mathcal{L}_{hw}$, and $\lambda$ is the coefficient applied to trade-off the task accuracy and efficiency. Note that for the shift and adder layers where the FLOP metric is not applicable, we first treat them as normal convolutional layers, and then scale the measured FLOPs down based on the computational cost of shift and adder layers normalized to that of corresponding multiplications to encourage hardware-friendly layers under the constraint of $\mathcal{L}_{hw}$.

However, the required search time and memory cost of vanilla DNAS grow linearly with the number of search candidates, challenging the exploration capability of our proposed NASA-NAS engine for the large hybrid search spaces defined in Table \ref{table:choice}. To tackle this limitation, we apply a masking mechanism to activate and optimize only several blocks within the supernet, so that the search cost is only determined by the number of active paths and thus agnostic to the total supernet size, inspired by \cite{ProxylessNAS}. Specifically, we activate blocks with top-$k$ sampling probabilities: 
\vspace{-0.6em}
\begin{equation}
	Y_{l} = \sum_{i = 1}^{n}GS(M(\alpha_{l}^{i}))w_{l}^{i}, \ l \in (0,22), \label{GS} \vspace{-0.6em}
\end{equation} 

where $Y_{l}$ is the output of searchable layer $l$ with $n$ candidate blocks, each of which is equipped with weights $w_{l}^{i}$ and architectural parameters $\alpha_{l}^{i}$ that represent the corresponding sampled probability. $M(\cdot)$ and $GS(\cdot)$ donate the masking and Gumbel Softmax (GS) \cite{GS} functions, respectively, which can be formulated as follows:
\vspace{-0.3em}
\begin{equation}
\begin{aligned}
	M(\alpha_{l}^{i})=\left\{
	\begin{array}{ll}
	\alpha_{l}^{i}       &    if\ {\alpha_{l}^{i} \in top_k(\alpha_{l})}\\
	0                    &    others   \\
	\end{array}\right.,
\\	GS(\alpha_{l}^{i}\vert \alpha_{l})=
	\frac{exp{(\alpha_{l}^{i}+g_{l}^{i})/\tau }}{\sum_{j = 1}^{n}exp{(\alpha_{l}^{j}+g_{l}^{j})/\tau}}, 
	\end{aligned} \vspace{-0.5em}
\end{equation} 
where $g_{l}^{i}$ is a random noise that obeys Gumbel(0, 1), and $\tau$ is a temperature parameter that controls the distribution of the GS function:  $\tau$ is first set as a large number to force the GS function to approximate a uniform distribution, and then gradually decreased for guiding the GS function towards a discrete sampling function.

After identifying the best architecture via our NASA-NAS engine, we train it from scratch to obtain its task accuracy.
\vspace{-0.5em}
\section{Proposed NASA-Accelerator Engine}
\label{NASA-Accelerator}
	\vspace{-0.2em}
In this section, we present our NASA-Accelerator aiming to better leverage the properties of NASA-NAS searched hybrid models for improved hardware efficiency. 
Specifically, Sec. \ref{sec:accelerator_chunk} introduces NASA-Accelerator's chunk-based architecture, which integrates dedicated sub-processors for heterogeneous hybrid layers. Then, we highlight the auto-mapper in Sec. \ref{sec:auto_mapper} that automatically schedules the searched hybrid models into dedicated accelerator to further improve the hardware efficiency.

\begin{figure}[!t]
	\centerline{\includegraphics[width=\linewidth]{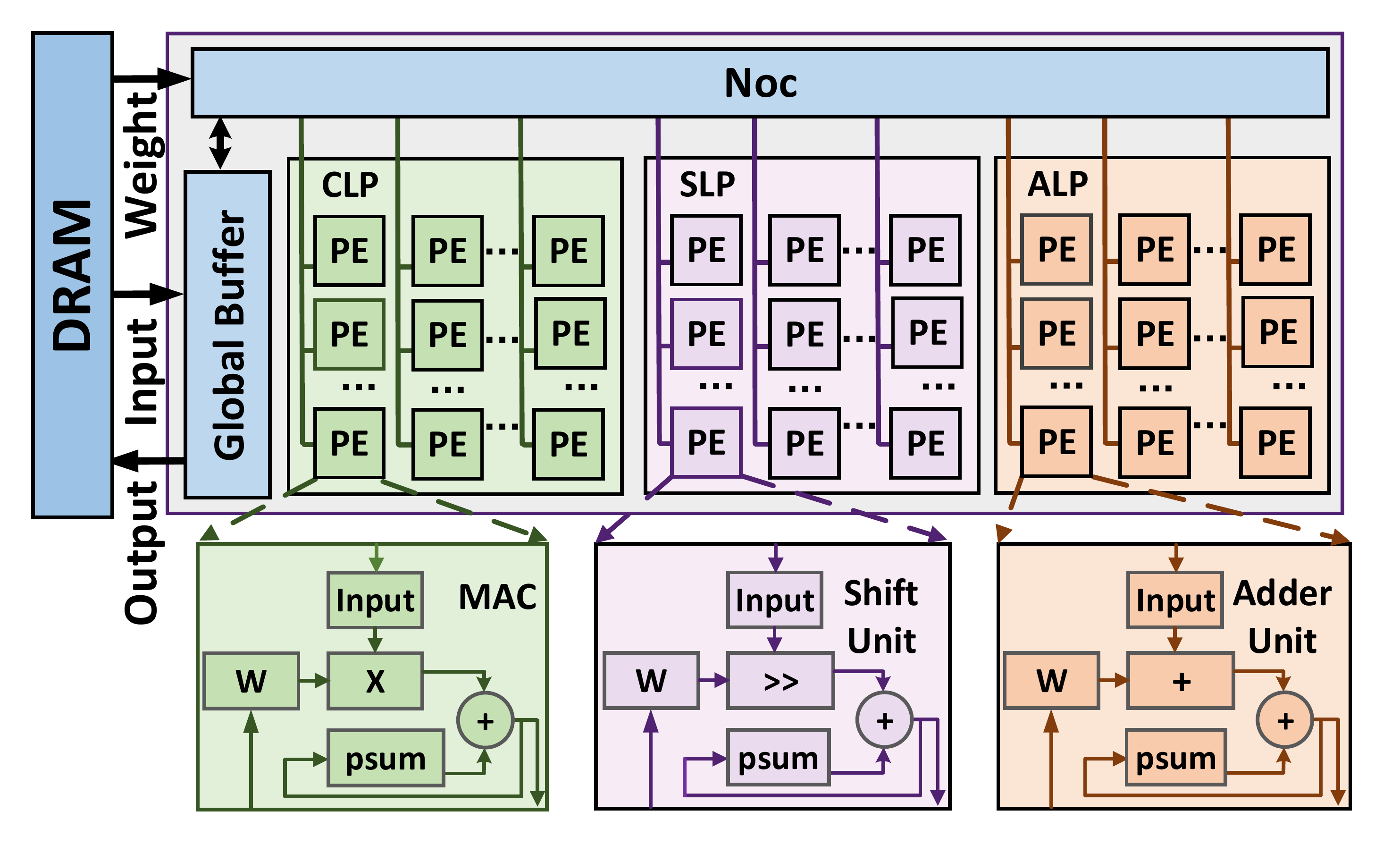}}
	\vspace{-1.5em}
	\caption{Micro-architecture of our NASA-Accelerator engine, which integrates dedicated sub-processors/chunks called CLP, SLP, and ALP to process the convolutional, shift, and adder layers in NASA-NAS searched hybrid models, respectively. A four-level memory hierarchy (i.e., DRAM, global buffer, NoC, and register files within each PE) is considered to facilitate data reuses, while customized PE units (i.e., MAC, Shift Units, and Adder Units) are designed to accelerate convolutional, shift, and adder layers.}
	\label{accelerator_overview} \vspace{-1.6em}
\end{figure}

\vspace{-0.5em}
\subsection{NASA-Accelerator's Micro-architecture}
\label{sec:accelerator_chunk}
\textbf{Architecture overview.}
Our NASA-Accelerator adopts (1) a multi-chunk micro-architecture to facilitate designing customized processing elements (PEs) for the heterogeneous layers in NASA-NAS searched hybrid models with (2) a four-level memory hierarchy to enhance data locality. Specifically, as shown in Fig. \ref{accelerator_overview}, our NASA-Accelerator architecture consists of an off-chip DRAM, an on-chip global buffer, a network on chip (NoC), and three sub-processors/chunks. Specifically, the global buffer caches the computational inputs and outputs on chip to reduce costly off-chip data accesses from DRAM, the NoC serves as a bridge between the global buffer and PE array for further enhanced data reuses and can read weights directly from the off-chip DRAM, and the three sub-processors denoting convolution/shift/adder layer processors (CLP/SLP/ALP) customize their PEs to better accelerate the convolutional, shift, and adder layers in hybrid models, respectively. In particular, multiplication and accumulation (MAC) units are utilized in CLP while bit-wise shift and accumulation units (Shift Units) and addition and accumulation units (Adder Units) are designed in SLP and ALP, respectively. Additionally, while the three sub-processors share the DRAM, global buffer, and NoC, each of their PE has its own register files (RFs) for storing inputs, weights, and partial sums, respectively. Overall, our NASA-Accelerator architecture aims to (1) tailor their PE designs for assigned algorithmic layers in NASA-NAS searched hybrid DNNs and (2) maximize data locality and thus the overall acceleration efficiency.

\begin{figure}[!t]
	\centerline{\includegraphics[width=0.9\linewidth]{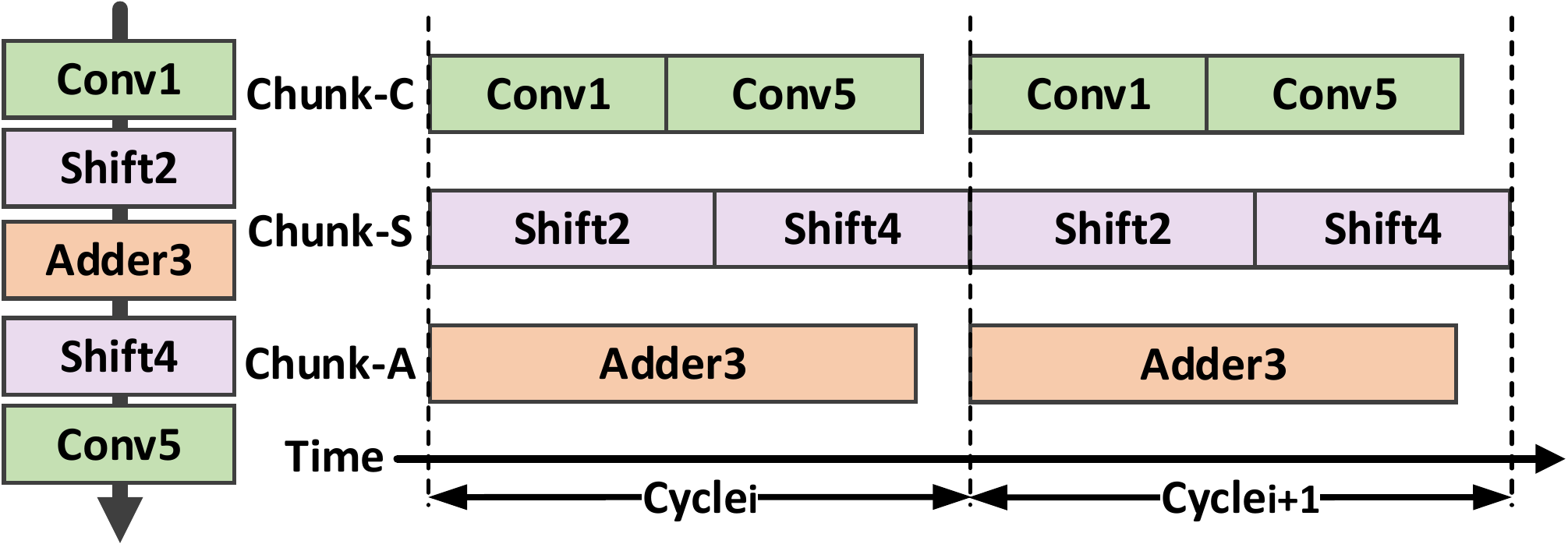}}
	\vspace{-1em}
	\caption{Temporal processing schedule of our chunk-based accelerator across cycles, in which each sub-accelerator sequentially processes assigned layers corresponding to independent data to optimize the hardware utilization.}
	\label{fig:pl_schedule} 
	\vspace{-1.8em}
\end{figure} 

\textbf{The PE allocation strategy.}
One important question for effectively designing the aforementioned multi-chunk micro-architecture is how to partition and then allocate the limited on-accelerator resources for the sub-processors dedicated for different types of layers in NASA-NAS searched hybrid models. For better understanding our allocation strategy, we first introduce the temporal processing schedule of our accelerator, where each sub-processor sequentially processes assigned layers with independent input data to ensure concurrent processing of all sub-processors for optimizing the overall hardware utilization. Fig. \ref{fig:pl_schedule} shows the temporal processing schedule for an NASA-NAS searched hybrid model with five layers as an illustrative example. During each cycle, sub-processors CLP, SLP, and ALP sequentially process the convolutional layers (Conv1 and Conv5), shift layers (Shift2 and Shift4), and adder layers (Adder3), respectively, each corresponds to different input data, e.g., the output of Conv1 processed by CLP in $cycle_{i}$ serves as input of Shift2 computed by SLP in $cycle_{i+1}$. From the above schedule and example, we can see that the achievable throughput of our accelerator is limited by the dominant latency across cycles. Hence, our PE allocation strategy strives to allocate PE resources to all three sub-processors for balancing their throughput and thus minimizing the latency of each cycle. 

Formally, our PE allocation strategy above can be formulated as:
\vspace{-0.5em}
\begin{equation}
    \begin{aligned}
        {N_{CLP}}/{O_{Conv}}={N_{SLP}}/{O_{Shift}}={N_{ALP}}/{O_{Adder}},\\
        \mathop{s.t.}\ A_{CLP} + A_{SLP} + A_{ALP} = Area\ Constraint.
        \end{aligned} \label{eq:pe-allcation} \vspace{-0.5em}
    \end{equation}
    
Simply, it ensures that the number of PEs allocated for sub-accelerators CLP ($N_{CLP}$), SLP ($N_{SLP}$), and ALP ($N_{ALP}$) are proportional to the total number of operations in the convolutional layers ($O_{Conv}$), shift layers ($O_{Shift}$), and adder layers ($O_{Adder}$), respectively, under the area budget. Note that the PEs of sub-accelerators SLP and ALP will be integrated jointly or separately with CLP to execute NASA-NAS searched hybrid models based on their layer types $T$ pre-defined in Table. \ref{table:choice}, under the same resource budget.  

Thanks to the area- and energy- efficient shift and adder layers \cite{ShiftAddNet}, our chunk-based accelerator dedicated for multiplication-reduced hybrid models can partially trade higher-cost MACs for convolutional layers with lower-cost Shift and/or Adder Units under the same area budget, increasing the parallelism and thus reducing the overall latency and energy cost.

\vspace{-0.8em}
\subsection{NASA-Accelerator's Auto-Mapper}
\label{sec:auto_mapper}
\vspace{-0.2em}
It is well-known that dataflows (i.e., temporal and spatial algorithm to accelerator mapping methods) can largely impact an accelerator's hardware efficiency, while it is challenging to figure out an optimal choice which requires jointly considering both the model structure and the hardware design. Considering that the larger dataflow space resulting from more heterogeneous layers in our target hybrid models, it is nontrivial and even more challenging to manually identify the optimal dataflow for a hybrid model to be processed in our accelerator. As such, our NASA framework integrates an automated dataflow search engine called auto-mapper to automatically search for optimal dataflows of a given hybrid model when being executed in our accelerator. 

In auto-mapper, we leverage the widely adopted \emph{nested for-loop description} \cite{EyerissAS,chen2018eyeriss,predictor} that are characterized by \emph{loop ordering factors} and \emph{loop tiling factors} to construct the dataflow search space. Specifically, loop ordering factors correspond to how to schedule the computations in the target PE array and within each PE, and thus determine the data reuse patterns, while loop tiling factors determine how to store data within each memory hierarchy to {effectively accommodate the above loop tiling factors}. 
Hence, the dataflow search space in our auto-mapper can be summarized as: 
    
\textbf{Loop ordering factors:} Determine the data reuse patterns. Here we search from four patterns: \emph{row stationary (RS), input stationary (IS), weight stationary (WS), and output stationary (OS)} for each chunk, and thus have a total of \emph{64 (4*4*4) choices} for reuse pattern of the three sub-accelerators in our accelerator.

\textbf{Loop tiling factors:} Determine how to store data within each memory hierarchy (e.g., DRAM, global buffer, NoC, and RF in our accelerator) to effectively accommodate the above data reuse patterns, and can be derived from \emph{all possible choices under the resource budget} (e.g., memory and computation resource budgets).

\vspace{-1.0em}
\section{Experimental Results}
\vspace{-0.2em}
In this section, we first describe the experiment setup in Sec. \ref{sec:set-up} and
the overall comparisons between our NASA framework and prior SOTA systems in Sec. \ref{sec:nasa-over-sota}, 
and then evaluate NASA's algorithm and hardware enablers, i.e., NASA-NAS and NASA-Accelerator, in 
Sec. \ref{sec:exp_alg} and Sec. \ref{sec:exp_hardware}, respectively.

\vspace{-1.2em}
\subsection{Experiment Setup}
\label{sec:set-up}
\vspace{-0.15em}
\textbf{Datasets, baselines and evaluation metrics.}
To verify the effectiveness of our proposed NASA framework, which integrates
(1) NASA-NAS to enable effective search for multiplication-reduced hybrid models;
and (2) NASA-Accelerator to boost hardware efficiency of NASA-NAS resulting models,
we consider \textbf{two datasets}: CIFAR10 and CIFAR100, and \textbf{three SOTA baselines}: 
(1) SOTA multiplication-based NAS work FBNet \cite{FBNet} on SOTA multiplication-based accelerator Eyeriss \cite{Eyeriss}; 
(2) SOTA handcrafted multiplication-free networks, i.e., DeepShift \cite{DeepShift} and AdderNet \cite{AdderNet}, on Eyeriss \cite{Eyeriss} with MACs in PEs replaced by the corresponding Shift and Adder Units (See Sec. \ref{sec:accelerator_chunk});
and (3) AdderNet \cite{AdderNet} on its dedicated accelerator \cite{AdderNetAI}.
We compare our NASA over the above SOTA systems in terms of \textbf{two evaluation metrics}: accuracy and Energy-Delay Product (EDP), a well-known hardware metric considering both energy and latency, \textit{under the same hardware budget for a fair comparison}. 

\textbf{Search and training recipes.}
We search for hardware-inspired hybrid models from the pre-defined three search spaces: \emph{hybrid-shift}, \emph{hybrid-adder}, and \emph{hybrid-all}.
\underline{For pretraining}, 
we pretrain the hybrid-shift supernet for $60$ epochs following the pretraining recipe as described in \cite{FBNet}. 
For hybrid-adder and hybrid-all supernets that include adder layers
and therefore suffer from the non-convergence problem,
we pretrain them with the proposed PGP strategy (See Sec. \ref{sec:progressive_training}).
Specifically, we pretrain the hybrid-adder supernet for $120$ epochs, while pretrain the hybrid-all supernet for $150$ epochs to accommodate the larger search space. 
\underline{For searching}, 
we start from the pretrained supernet and iteratively train the model weights and architecture parameters for $90$ epochs with a batch size of $128$, following \cite{FBNet}. 
In each epoch, the weights are trained on $50\%$ of the training dataset using the SGD optimizer with a momentum of $0.9$. The initial learning rate (lr) is $0.05$ for hybrid-shift and $0.1$ for other search spaces, and will be gradually decayed following a cosine scheduler. 
The architecture parameters are optimized on the rest $50\%$ of the training set by Adam optimizer \cite{Adam} 
with a learning rate of 3e-4 and weight decay of 5e-4.
The initial temperature $\tau$ of Gumbel Softmax is set to $5$ and decayed by $0.956$ every epoch following \cite{FBNet}. 
\underline{For training}, 
we train the searched networks from scratch for $650$ epochs with a batch size of $196$, an initial lr of $0.02$ with a cosine lr scheduler for hybrid-shift models, and an initial lr of $0.1$ with a multi-step scheduler for hybrid-adder and hybrid-all models.

\textbf{Hardware experiment setup.}
To verify the performance of NASA-Accelerator, we implement a cycle-accurate simulator on top of \cite{predictor} as our chunk-based accelerator to obtain fast and reliable estimations.
We verified it against the RTL implementation to ensure its correctness.
The adopted unit energy and area are synthesized on CMOS $45nm$ technology at a frequency of $250MHz$.
For fair comparisons, we also follow \cite{predictor} to implement cycle-accurate simulators as benchmark accelerators: 
Eyeriss equipped with MACs for multiplication-based FBNet \cite{FBNet}, equipped with Shift Units for multiplication-free DeepShift \cite{DeepShift}, and equipped with Adder Units for AdderNet \cite{AdderNet}.
We evaluate both NASA-Accelerator and the above baselines under the same hardware constraint, CMOS technology, and clock frequency. 
Notably, to execute our hybrid models and other baselines into corresponding hardware accelerators, we follow SOTA quantization method \cite{ScalableMF} to quantize both their weights and activations into $8$ bits, except for those in shift and adder layers of our hybrid models, which are quantized into $6$ bit.

\vspace{-1em}
\subsection{NASA over SOTA systems}
\label{sec:nasa-over-sota}

\begin{figure}[htbp]
	\centerline{\includegraphics[width=\linewidth]{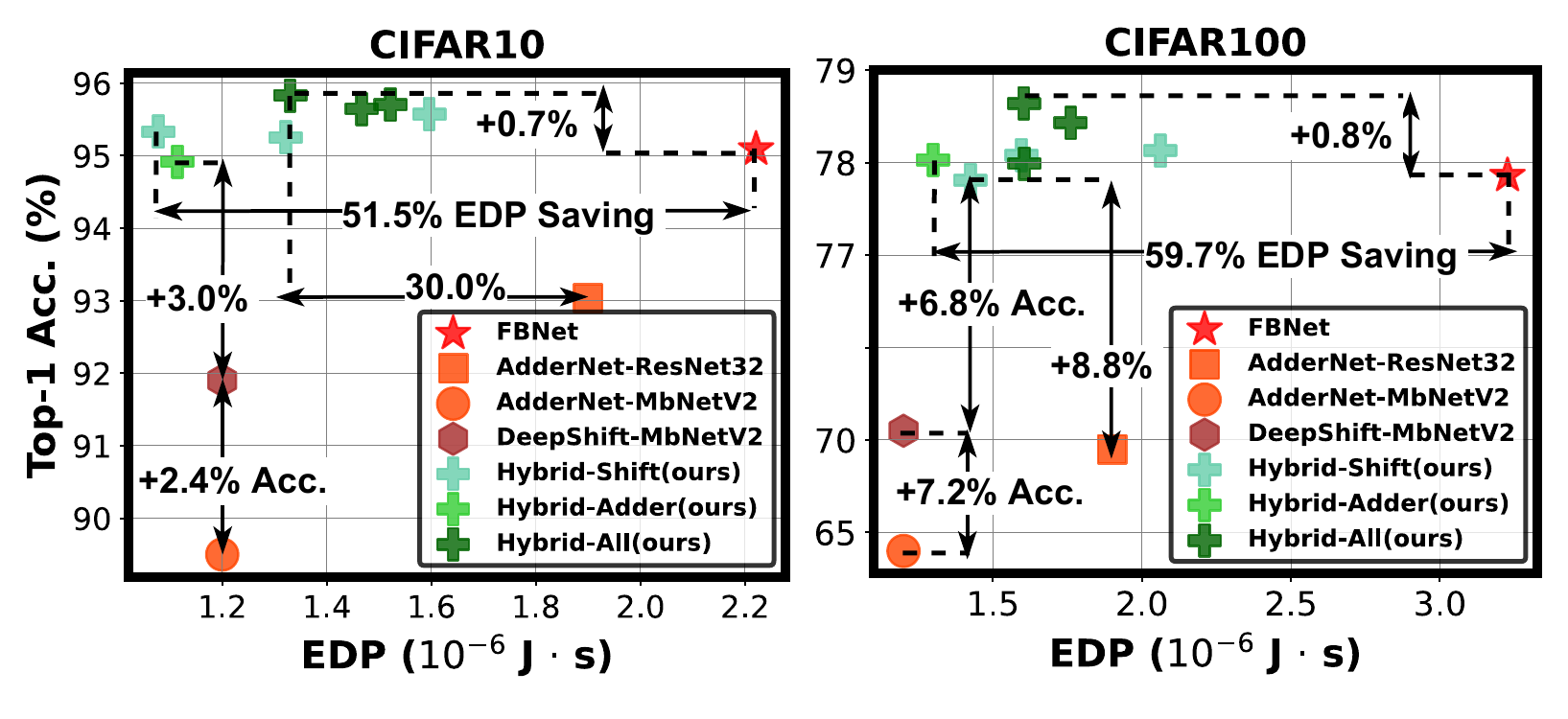}}
	\vspace{-1.7em}
	\caption{Benchmarking NASA over SOTA multiplication-based and multiplication-free baselines on CIFAR10 and CIFAR100 in terms of accuracy and EDP.} 
	\label{fig:over-sota} \vspace{-1.7em} 
\end{figure} 
\vspace{-0.3em}
As shown in Fig. \ref{fig:over-sota}, 
the proposed NASA framework
consistently surpasses all SOTA baselines, e.g., SOTA multiplication-based and multiplication-free models on Eyeriss with merely MACs and Shift/Adder Units (as introduced in Sec. \ref{sec:accelerator_chunk}) in terms of accuracy and EDP trade-off.
Note that we compare under the same area and memory budget for fair.
Specifically, 
(1) NASA achieves much higher accuracy over SOTA multiplication-free systems under comparable or even better hardware efficiency, thanks to the powerful convolutions in NASA-NAS resulting hybrid-models and the high-parallelism chunk-based design in NASA-Accelerator.
In particular, NASA achieves up to \textbf{6.8\%} and \textbf{14.0\%} accuracy improvements over DeepShift-MobileNetV2 and AdderNet-MobileNetV2 on Eyeriss with customized multiplication-free units under the same hardware resource budget, with almost the same EDP (\textbf{$\pm$-0.1\%}).
Moreover, NASA offers \textbf{8.8\%} higher accuracy with \textbf{25.0\%} EDP saving against AdderNet-ResNet32 on \cite{AdderNetAI}, when testing on CIFAR100.
(2) NASA also outperforms the SOTA multiplication-based system, i.e., SOTA NAS work FBNet \cite{FBNet} on Eyeriss \cite{Eyeriss} with MACs, by fully leveraging the algorithmic benefit of multiplication-reduced searched models and thus offering comparable or even higher accuracy with much lower EDP, e.g., we can achieve up to 0.24\% and 0.20\% higher accuracy with \textbf{51.5\%} and \textbf{59.7\%} lower EDP on CIFAR10 and CIFAR100, respectively.
This set of experiments help to verify the effectiveness of NASA algorithm and hardware co-design system for boosting both the accuracy and hardware efficiency. 

\vspace{-0.8em}
\subsection{Evaluation of NASA-NAS}
\label{sec:exp_alg}
\vspace{-0.2em}

\begin{table*}[]
\centering
\caption{The operation numbers (e.g. multiplications, bit-wise shifts, and additions) and accuracy comparisons of NASA-NAS resulting hybrid models over SOTA multiplication-free and multiplication-based models on CIFAR10 and CIFAR100.} \vspace{-1em}
{
\resizebox{\linewidth}{!}{
\begin{tabular}{c|ccc|cc|ccc|cc}
\hline \hline
                                           & \multicolumn{5}{c|}{CIFAR10} & \multicolumn{5}{c}{CIFAR100}     \\ \cline{2-11}
                                           & \multicolumn{3}{c|}{Operation Numbers} & \multicolumn{2}{c|}{Acc.(\%)}     & \multicolumn{3}{c|}{Operation Numbers} & \multicolumn{2}{c}{Acc.(\%)}     \\ \cline{2-11} 
\multirow{-3}{*}{Models}                   & Mult.      & Shift     & Addition     & FP32                & FXP8             & Mult.      & Shift     & Addition     & FP32                & FXP8             \\ \hline
DeepShift-MobileNetV2\cite{DeepShift}                        & 3.3M       & 39.6M     & 42.9M        & --                  & 91.9(-3.2)          & 3.35M      & 39.6M     & 42.9M        & --                  & 71.0(-7.2)          \\ 
AdderNet-MobileNetV2\cite{AdderNet} & 3.3M       & 0M        & 82.5M        & 90.5(-4.9)          & 89.5(-5.6)          & 3.35M      & 0M         & 82.5M        & 64.1(-14.1)         & 63.5(-14.4)         \\ \hline
FBNet\cite{FBNet}                                      & 47.2M      & 0M        & 47.2M        & 95.4                & 95.1                & 56.6M      & 0M         & 56.6M        & 78.2                & 77.9                \\ \hline
Hybrid-Shift-A                             & 29.6M      & 21.6M     & 51.2M        & \textbf{95.5(+0.1)} & \textbf{95.6(+0.5)} & 34.3M      & 24.1M     & 58.5M        & 78.2(+0.0)          & \textbf{78.2(+0.3)} \\
Hybrid-Shift-B                             & 32.3M      & 11.1M     & 43.5M        & \textbf{95.5(+0.1)} & \textbf{95.3(+0.2)} & 35.1M      & 23.6M     & 58.8M        & \textbf{78.3(+0.1)} & \textbf{78.1(+0.2)} \\
Hybrid-Shift-C                             & 24.6M      & 18.1M     & 42.7M        & 95.3(-0.1)          & \textbf{95.3(+0.2)} & 33.9M      & 14.6M     & 48.6M        & 78.0(-0.2)          & 77.8(-0.1)          \\ 
Hybrid-Adder-A                             & 35.4M      & 0M        & 43.8M        & 95.0(-0.4)          & 94.9(-0.2)          & 29.9M      & 0M         & 60.1M        & \textbf{78.4(+0.2)} & \textbf{78.1(+0.2)} \\ 
Hybrid-All-A                               & 24.1M      & 23.8M     & 63.5M        & \textbf{95.7(+0.3)} & \textbf{95.7(+0.6)} & 27.5M      & 19.9M     & 64.7M        & \textbf{78.4(+0.2)} & \textbf{78.0(+0.1)} \\
Hybrid-All-B                               & 27.7M      & 17.2M     & 68.5M        & \textbf{95.9(+0.5)} & \textbf{95.7(+0.6)} & 25.2M      & 20.5M     & 64.5M        & \textbf{78.7(+0.5)} & \textbf{78.7(+0.8)} \\
Hybrid-All-C                               & 21.0M      & 20.3M     & 64.6M        & \textbf{95.8(+0.4)} & \textbf{95.8(+0.7)} & 28.2M      & 23.9M     & 68.1M        & \textbf{78.3(+0.1)} & \textbf{78.5(+0.6)} \\ \hline \hline
\end{tabular}
\vspace{-1.5em}
}} \label{table:NASA-results} 
\end{table*}

\textbf{Results and analysis.}
We compare the NASA-NAS resulting multiplication-reduced models with SOTA handcrafted multiplication-free networks, including DeepShift-MobileNetV2 \cite{DeepShift} and AdderNet-MobileNetV2 \cite{AdderNet}, and SOTA multiplication-based models searched by FBNet \cite{FBNet}, on both CIFAR10 and CIFAR100. 
As shown in Table \ref{table:NASA-results}, we have three observations: 
(1) our hybrid models consistently outperform all baselines in terms of accuracy-efficiency trade-offs regardless of full precision (FP32) or 8-bit fixed-point (FXP8) settings, verifying NASA-NAS's effectiveness and the hypothesis that we can unify powerful convolutions with energy-efficient shift and adder layers to build hybrid models for boosting task accuracy;
(2) NASA achieves on-average 0.13\% and 0.34\% accuracy improvements over full precision and FXP8 multiplication-based baselines, respectively, benefit from the hardware friendly operators such as shift and adder, \emph{demonstrating that our hybrid models are powerful and robust to quantization};
and (3) as for the comparison within the three search spaces pre-defined in NASA, hybrid-all models consistently outperform other searched models by achieving a higher accuracy, showing the superiority of hybrid-all search space that includes both coarse-grained shift and fine-grained adder layers.


\begin{figure}[t]
	\centerline{\includegraphics[width=0.9\linewidth]{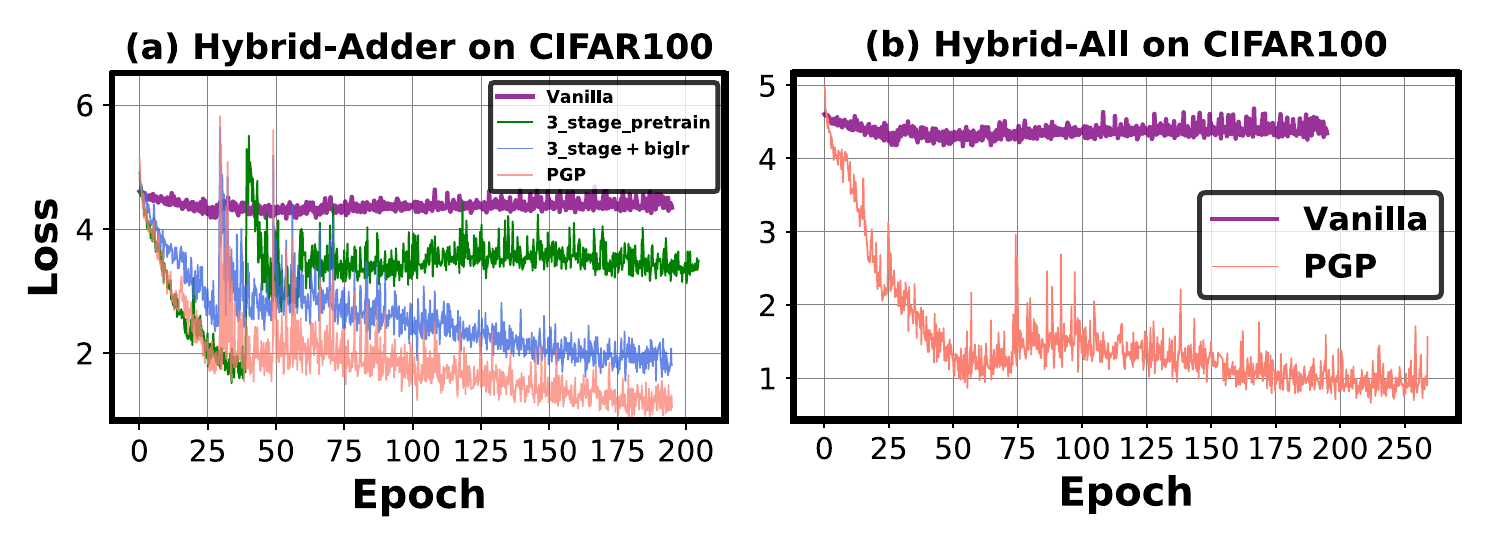}}
	\vspace{-1.8em}
	\caption{Ablation study of the progressive pretrain strategy on (a) hybrid-adder and (b) hybrid-all search spaces.} 
	\label{fig:pgp} 
	\vspace{-1.5em}
\end{figure} 
\begin{figure}[t]
	\centerline{\includegraphics[width=0.9\linewidth]{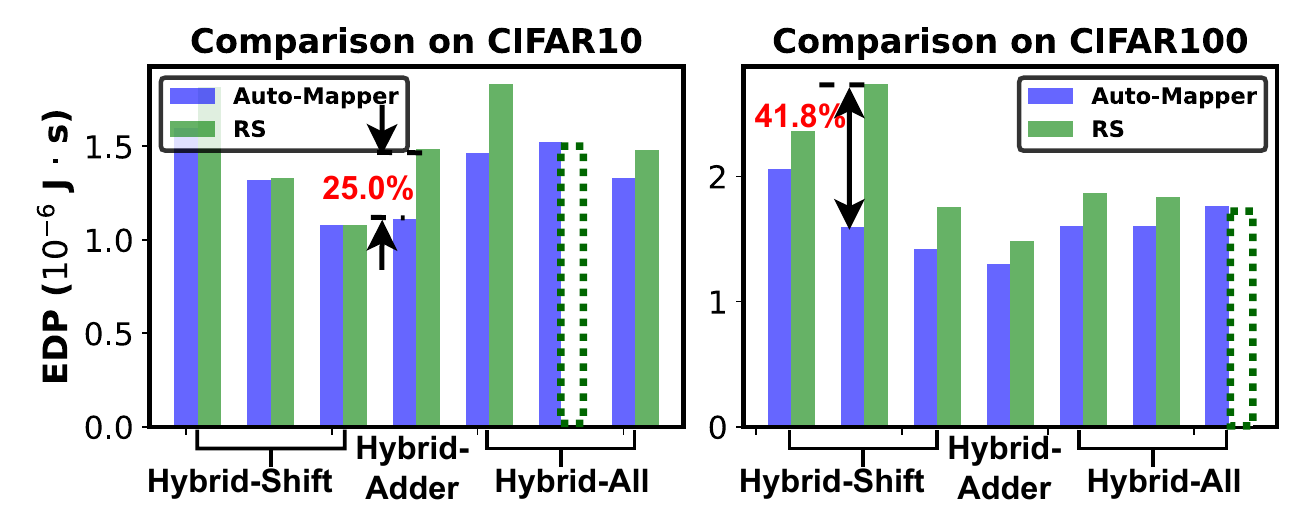}}
	\vspace{-1.3em}
	\caption{Auto-Mapper over SOTA expert-crafted RS.}  
	\label{fig:auto-mapper}
	\vspace{-2.1em}
\end{figure} 
\textbf{{Ablation study of PGP.}} 
Fig. \ref{fig:pgp} (a) visualizes the training trajectories, from which we can see that: (1) the mismatch between adder layers and convolutions in terms of weight distribution and convergence speed yields the non-convergence problem when searching for hybrid-adder models, while the vanilla pretrain method in \cite{FBNet} can not handle this issue, calling for customized training strategy; 
(2) our proposed PGP accelerates the convergence during search, verifying PGP’s effectiveness and our hypothesis that progressively training heterogeneous layers can promote the fusion of them; 
(3) the three-stage pretrain method in PGP sets a good initialization for hybrid-adder supernet via training convolutions and adder layers sequentially and then jointly, and the big lr alleviates the slow convergence problem of adder layers and thus accelerates the optimization process.
Besides, the initialization that proposed to stabilize the search process in \cite{BigNAS} also benefits the training of our hybrid supernet. 
We observe the same non-convergence problem when training the hybrid-all supernet, and PGP also works well as proofed in Fig. \ref{fig:pgp} (b). 
\vspace{-1em}

\subsection{Evaluation of the NASA-Accelerator}
\label{sec:exp_hardware}
\vspace{-0.3em}
\textbf{Effectiveness of auto-mapper.}
Fig. \ref{fig:auto-mapper} shows the comparison between Row Stationary (RS) mapping and our proposed auto-mapper, from which we can observe that: 
(1) the heterogeneous dataflows suggested by auto-mapper consistently outperform the SOTA expert-crafted dataflow, e.g., RS for each chunk which is also included in our search space, verifying the auto-mapper's effectiveness; 
(2) in particular, auto-mapper can achieve up to \textbf{25.0\%} and \textbf{41.8\%} EDP saving on CIFAR10 and CIFAR100, respectively; 
and (3) it is worth noted that due to the competition between chunks, e.g., CLP, SLP, and ALP that are executed to process corresponding convolutions, shift, and adder layers in parallel, for hardware resources such as shared buffer, the fixed RS for all chunks fails to map hybrid models into the chunk-based accelerator under given hardware constraint in some cases (e.g., those indicated in green dotted line).
\vspace{-1.3em}

\section{Conclusion}
\vspace{-0.3em}
We propose, develop, and validate NASA, \emph{the first} algorithm and hardware co-design framework dedicated for hybrid DNN models. Our NASA framework integrates a NAS and accelerator engine customized for developing and accelerating hybrid DNN models, respectively. Specifically, 
NASA-NAS fuses hardware friendly operators, such as bit-wise shifts and additions, into a SOTA FBNet search space, and equips a NAS algorithm with our proposed progressive pretrain strategy to identify optimal hybrid DNNs from our constructed hybrid search spaces; NASA-Accelerator advocates a chunk-based accelerator and integrates an auto-mapper to optimize dataflows for heterogeneous layers in hybrid DNNs. Extensive experimental results validate the effectiveness and advantages of our NASA framework in boosting DNN accuracy and efficiency. We believe our work can open up an exiting perspective for the exploration and deployment of multiplication-reduced hybrid models to boost both task accuracy and efficiency.

\bibliographystyle{ACM-Reference-Format}

\end{document}